\documentclass[12pt]{amsart}

\newcommand {\supplus}{\mathop{{\supset}\llap{\raise 
0.5pt\hbox{\normalfont\small+}\hskip 0.5pt}}} 

\newcommand {\subplus}{\mathop{{\subset}\llap{\raise 
0.5pt\hbox{\normalfont\small+}\hskip 0.5pt}}}  

\newcommand {\cal} {\mathcal}

\def \opname#1#2%
  {\expandafter\newcommand \csname #1\endcsname {{\mathop{#2}\nolimits}}}

\newcommand{\rmname}[1]
  {\expandafter\newcommand \csname #1\endcsname {{\operatorname{#1}}}}

\newcommand{\rmnameii}[2]
  {\expandafter\newcommand \csname #1\endcsname {{\operatorname{#2}}}}

\rmname{act}
\rmname{Ad}
\rmname{Add}
\rmname{ad}
\rmname{Alt}
\rmname{alt}
\rmname{Ann}
\rmname{antidiag}
\rmname{Ber}
\rmname{ber}
\rmname{Br}
\rmname{card}
\rmname{ch}
\rmname{Char}
\rmname{cem}
\rmname{cj}
\rmname{Cliff}
\rmname{cntr}
\rmname{codim}
\rmname{coind}
\rmname{const}
\rmname{col}
\rmname{cork}
\rmname{cpr}
\rmname{diag}
\rmnameii{Div}{div}
\rmname{Def}
\rmname{Der}
\rmname{Diff}
\rmname{Dim}
\rmname{End}
\rmname{Even}
\rmname{Ext}
\rmname{gr}
\rmname{Hom}
\rmname{HT}
\rmnameii{Ht}{ht}
\rmname{hwt}
\rmname{Id}
\rmname{id}
\rmname{ind}
\rmname{Ind}
\rmname{Inf}
\rmname{irr}
\rmname{Le}
\rmname{Lie}
\rmname{lwt}
\rmname{mult}
\rmname{Mat}
\rmname{Mor}
\rmname{nm}
\rmname{Ob}
\rmname{Odd}
\rmname{Osc}
\rmname{per}
\rmname{Pic}
\rmname{pr}
\rmname{pro}
\rmname{Prime}
\rmname{Proj}
\rmname{prt}
\rmname{pt}
\rmname{Q}
\rmname{qet}
\rmname{qtr}
\rmname{rd}
\rmname{rk}
\rmname{row}
\rmname{Res}
\rmname{salt}
\rmname{Sch}
\rmname{SBr}
\rmname{scalar}
\rmname{Ser}
\rmname{sign}
\rmname{Smbl}
\rmname{spin}
\rmname{ssym}
\rmname{str}
\rmname{st}
\rmname{sgn}
\rmname{sq}
\rmname{symm}
\rmname{supp}
\rmname{Supp}
\rmname{St}
\rmname{Spec}
\rmname{Spm}
\rmname{tr}
\rmname{vpt}
\rmname{weyl}
\rmname{Weyl}
\rmname{Witt}

\opname{vvol}  {{v\hspace{-0.1ex}o\hspace{-0.02ex}l\/}}
\opname{pnt}  {\text{\normalfont pt}}
\opname{Span} {{Span}}
\opname{slim} {\overline{\lim}}
\opname{Vol}  {{V\hspace{-0.55ex}o\hspace{-0.02ex}l\/}}
\opname{Par}  {{P\hspace{-0.3ex}a\hspace{-0.05ex}r\/}}

\newcommand {\bcdot}   {\mathbin{\hbox{\raise.4ex\hbox{\bf.}}}} 

\newcommand {\secno} {}

\newtheorem{Theorem}{\secno Theorem}

\newenvironment {th*}[1]
    {\gdef\thname{#1} \begin{thn}}%
    {\end{thn}}
\newtheorem{thn}[Theorem] {\thname}

\theoremstyle{definition}

\newenvironment {ex*}[1]
    {\gdef\thname{#1} \begin{exn}}%
    {\end{exn}}
\newtheorem{exn}[Theorem]{\thname}

\theoremstyle{remark}

\newenvironment {rem*}[1]
    {\gdef\thname{#1} \begin{remn}}%
    {\end{remn}}
\newtheorem{remn}[Theorem]{\thname}

\begin{document}

\title{An unconventional supergravity}

\author{Pavel Grozman, Dimitry Leites}

\address{Department of Mathematics, University of Stockholm,
Kr\"aftriket hus 6, SE-106 91, Stockholm, Sweden}

\thanks{We gratefully acknowledge partial financial support of The
Swedish Institute and an NFR grant, respectively.}

\begin{abstract} We introduce and completely describe the analogues of
the Riemann curvature tensor for the curved supergrassmannian of the
passing through the origin $(0|2)$-dimensional subsupermanifolds in
the $(0|4)$-dimensional supermanifold with the preserved volume form. 
The underlying manifold of this supergrassmannian is the conventional
Penrose's complexified and compactified version of the Minkowski
space, i.e., the Grassmannian of $2$-dimensional subspaces in the
$4$-dimensional space. 

The result provides with yet another counterexample to
Coleman--Mandula's theorem.
\end{abstract} 

\maketitle

This paper should appear in: S.~Duplij and J.~Wess (eds.)  {\em
Noncommutative structures in mathematics and physics}, Proc.  NATO
Advanced Reserch Workshop, Kiev, 2000.  Kluwer, 13--30.

{\bf 1.  New supertwistors}.  Penrose suggested an unusual description
of our space-time, namely, to compactify the Minkowski space-time
model of the Universe (nontrivially: with a light cone at the
infinity) and complexify this compactification.  The final result is
$Gr_{2}^4$, the Grassmanian of 2-dimensional subspaces in the
4-dimensional (complex) space (of so-called twistors).  There are many
papers and several monographs on advantages of this interpretation of
the space-time in various problems of mathematical physics; we refer
the reader to Manin's book \cite{Man}, where an original Witten's idea
to incorporate supervarieties and consider infinitesimal neighborhoods
for interpretation of the ``usual'', i.e., non-super, Yang-Mills
equations is thouroghly investigated together with several ways to
superize Minkowski space.  Ours is one more, distinct, way.

Observe that the supermanifold of $(0|2)$-dimensional subsuperspaces
in the $(0|4)$-dimensional superspace is identical with $Gr_{2}^4$,
only the tautological bundle is different: the fiber is purely odd. 
In this work we consider not subsuper{\it spaces} but subsuper{\it
manifolds}.

We considered the structure functions --- analogs of the Riemann
tensor --- for the {\it curved supergrassmannian} ${\mathcal
CGr}_{0|2}^{0|4}$ of $(0|2)$-dimensional subsuper{\it manifolds} in
the $(0|4)$-dimensional supermanifold.  Recall that the ``usual''
grassmannian consists of linear subspaces of the linear space passing
through the origin whereas the curved one consists of submanifolds, in
other words, nonlinear embeddings are allowed and the submanifolds do
not {\it have} to pass through a fixed point.  Obviously, the curved
Grassmannian is infinite dimensional, but the curved {\it
super}grassmannian ${\mathcal CGr}_{0, k}^{0, n}$ is of finite
superdimension: it is a quotient of the supergroup of
superdiffeomorphisms of the linear supermanifold ${\mathcal C}^{0, n}$
(the Lie superalgebra of this Lie supergroup is
$\mathfrak{vect}(0|n)=\mathfrak{der}\, {\mathbb C}[\theta_{1}, \dots,
\theta_{n}]$).  For the list of classical superspaces including curved
supergrassmannians see \cite{Lei/Ser/Vin}.)  The underlying manifold
of ${\mathcal CGr}_{0|2}^{0|4}$ is the conventional $Gr_{2}^4$ but
${\mathcal CGr}_{0, 2}^{0, 4}$ has also odd coordinates.

On ${\mathcal CGr}_{0|2}^{0|4}$, we have expanded the curvature
supertensor in components with respect to the (complexification of
the) Lorentz group and saw that it does not contain the components
used for the ordinary Einstein equations (EE), namely, there is no Ricci
curvature ${\rm Ric}$ and no scalar curvature ${\rm Scalar}$ (in what
follows $R(22)$ and $R(00)$, respectively).

So we decided to amend the initial model and consider the
supergrassmanian ${\mathcal CGr}_{0|2}^{0|4}(0)$ of subsupermanifolds
through the origin.  It turns out that this does not help, either: no ${\rm
Ric}$ and ${\rm Scalar}$.

We decided not to give up, and took for the model of Minkowski
superspace the supergrassmannian ${\mathcal SCGr}_{0|2}^{0|4}(0)$ of
subsupermanifolds through the origin with the volume element of the
ambient and the subsupermanifolds preserved.  On ${\mathcal
SCGr}_{0|2}^{0|4}(0)$, the expansion of the curvature supertensor does
contain $R(22)$ and $R(00)$!  There are no analogs of conformal (off
shell) structure functions.

Our model and its supergroup of motion --- an analogue of the
Poincar\'e group --- do not contradict the restrictions of the famous
no-go theorems by Haag--\L opuszanski--Sohnius and Coleman--Mandula
(for further discussions see \cite{del}) and provides us with a new,
missed so far, version of the Poincar\'e supergroup. The analogues
of Einstein equations we suggest are a totally new version of SUGRA.
Equating to zero other conformally non-invariant components we get
extra conditions; we do not know how to interprete them.

We do not see any reason for discarding this and similar models.  In
particular, we suggest to analyze the structurre functions (definition
below) on ${\mathcal CGr}_{0|2}^{0|4}$ and ${\mathcal
CGr}_{0|2}^{0|4}(0)$ which we have abandoned above.

The conventional reading of Coleman--Mandula's theorem (cf. 
\cite{Sal/Sez}) assumes that the complexified Lorentz Lie algebra
$\mathfrak{L}=\mathfrak{sl}(2)_L\oplus \mathfrak{sl}(2)_R$ {\it
commutes} with the Lie algebra of internal symmetries $\mathfrak{i}$
(for us $\mathfrak{i}$ is equal to $\mathfrak{sl}(2)_L\otimes{\mathbb C}
\xi_1\xi_2$, see sec.  4).

In our case $\mathfrak{L}$ acts on $\mathfrak{i}$ and forms a
semidirect sum with it; the bracket on $\mathfrak{i}$ is identically
zero.  This possibility does not contradict assumptions of
Coleman--Mandula's theorem but was not considered.

The odd parameters have a correct statistics with respect to the 
Lorentz Lie algebra.

We represent Einstein's equations as conditions on conformally
noninvariant components of the analog of the Riemann tensor, and
represent the Riemann tensor as a section of the bundle on the
(locally) Minkowski space whose fiber is certain {\bf Lie algebra
cohomology}.  This is a more user-friendly description of the
Riemannian tensor than the classical treatment of obstructions to
nonflatness in differential geometry.  We have in mind {\it Spencer
homology}, cf.  \cite{Ste}, where the case of any $G$-structure, not
only $G=O(n)$ is considered.  Superization of the definitions from
\cite{Ste} is the routine straightforward application of the Sign
Rule.

{\bf Remark}.  It is interesting to test the whole list of curved
supergrassmannians with the simple Lie supergroup of motion (see
Tables in \cite{Lei/Ser/Vin}) and similarly to the above sacrify the
simplicity of the supergroup of motion in order to get EE. Grozman's
package SuperLie (see \cite{Gro/Lei1}) is a useful tool in this
research problem: without a computer (and a good code) this task is
hardly feasible.

{\bf 2. Structure functions: recapitulation (\cite{Ste})}. Let $F(M)$ be 
the frame bundle over a manifold $M$, i.e., the principal 
$GL(n)$-bundle.  Let $G\subset GL(n)$ be a Lie group.  A $G$-{\it 
structure on} $M$ is a reduction of the principal $GL(n)$-bundle to 
the principal $G$-bundle.

The simplest $G$-structure is the {\it flat} $G$-structure defined as
follows.  Let $V$ be ${\mathbb R}^n$ (or ${\mathbb C}^n$) with a fixed
frame.  The flat structure is the bundle over $V$ whose fiber over
$v\in V$ consists of all frames obtained from the fixed one under the
$G$-action, $V$ being identified with $T_vV$ by means of the parallel
translation by $v$.

{\bf Examples of flat structures}.  The classical spaces, e.g., 
compact Hermitian symmetric spaces, provide us with 
examples of manifolds with nontrivial topology but flat $G$-structure.

In \cite{Ste} the obstructions to identification of the $k$th 
infinitesimal neighbourhood of a point $m$ on a manifold $M$ with 
$G$-structure with the $k$th infinitesimal neighbourhood of a point 
of the flat manifold $V$ with the above described flat $G$-structure 
are called {\it structure functions of order} $k$.  In \cite{Ste} it is 
shown further that the tensors that constitute these obstructions are 
well-defined {\it provided} the structure functions of all orders $<k$ 
vanish.  (In supergravity the conditions that structure functions of 
lesser orders vanish are called {\it Wess-Zumino constraints}.)

The classical description of the structure functions uses the notion
of the {\it Spencer cochain complex}.  Let us recall it.  Let $S^i$
denote the operator of the $i$-th symmetric power.  Set
$\mathfrak{g}_{-1} = T_mM$, let $\mathfrak{g}_0$ be the Lie algebra of
$G$; for $i > 0$ set:
$$
\begin{array}{c}
\mathfrak{g}_i = \{X\in {\rm Hom}~(\mathfrak{g}_{-1}, 
\mathfrak{g}_{i-1})\mid X(v_{0})(v_{1}, \dots, v_{i}) = 
X(v_{1})(v_{0}, 
\dots , v_{i})\; \cr
{\rm  for  \; \; any}\; \; \; v_{0}, v_{1}, \dots, v_{i}\in 
\mathfrak{g}_{-1}\}. \end{array}\eqno{(2.1)}
$$
Finally, set $(\mathfrak{g}_{-1}, \mathfrak{g}_0)_* = \mathop{\oplus}
\limits _{i\geq -1}\mathfrak{g}_i$.  This is the Lie algebra of {\it
all} transformations that preserve on $\mathfrak{g}_{-1}$ the same
structure which is preserved by the linear transformations from
$\mathfrak{g}_0$.

Suppose that the $\mathfrak{g}_0$-module $\mathfrak{g}_{-1}$ is
faithful, i.e., each nonzero element from $\mathfrak{g}_0$ acts
nontrivially.  Then, clearly, 
$$
(\mathfrak{g} _{-1},
\mathfrak{g}_0)_*\subset \mathfrak{vect}(n) = \mathfrak{der}\,
{\mathbb R} ~[x_1, ...  , x_n],
$$ 
where $n =\dim ~\mathfrak{g}_{-1}$,
with 
$$
\mathfrak{g}_i=\{X\in\mathfrak{vect}(n)_i: [X,
D]\in\mathfrak{g}_{i-1}{\rm  for \; \; any } D\in\mathfrak{g}_{-1}\}
$$ 
for $i\geq 1$.  It is easy to check that $(\mathfrak{g}_{-1},
\mathfrak{g}_0)_*$ is a Lie subalgebra of $\mathfrak{vect}(n)$.

The Lie algebra $(\mathfrak{g}_{-1}, \mathfrak{g}_0)_*$ will be called
the {\it Cartan's prolong} (the result of {\it Cartan's prolongation})
of the pair $(\mathfrak{g}_{-1}, \mathfrak{g}_0)$.

Let $E^i$ be the operator of the $i$-th exterior power; set (prime denotes dualization)
$$
C^{k, s}_{(\mathfrak{g}_{-1}, \mathfrak{g}_0)} =
\mathfrak{g}_{k-s}\otimes E^s(\mathfrak{g}_{-1}').
$$ 

Define the differential $\partial _s: C^{k, s}_{(\mathfrak{g}_{-1},
\mathfrak{g}_0)} \longrightarrow C^{k, s+1}_{(\mathfrak{g}_{-1},
\mathfrak{g}_0)} $ by setting for any $v_1, \dots, v_{s+1}\in V$ (as
usual, the slot with the hatted variable is to be ignored):
$$
(\partial _sf)(v_1, \dots , v_{s+1}) = \sum (-1)^i [f(v_1, \dots ,
\hat v _{s+1-i}, \dots , v_{s+1}),  v_{s+1-i}].\eqno{(2.2)}
$$
As expected, $\partial _s\partial _{s+1} = 0$, and the homology $H^{k,
s}_{(\mathfrak{g}_{-1}, \mathfrak{g}_0)} $ of the bicomplex
$\mathop{\oplus} \limits _{k, s}C^{k,s}_{(\mathfrak{g}_{-1},
\mathfrak{g}_0)} $ is called the $(k, s)$-th {\it Spencer cohomology}
of $(\mathfrak{g} _{-1}, \mathfrak{g}_0)_{*}$.  (Observe that we use a
grading of the Spencer complex different form that in \cite{Ste}. 
Ours is a more natural one.)

{\bf Proposition} (\cite{Ste}) {\sl The structure functions of order 
$k$ constitute the space of the $(k, 2)$-th Spencer cohomology of the 
$(\mathfrak{g}_{-1}, \mathfrak{g}_0)_{*}$.
}

{\bf 3.  Spencer cohomology in terms of Lie algebra cohomology}. We 
observe that
$$
\mathop{\oplus}\limits_k H^{k, 2}_{(\mathfrak{g}_{-1}, \mathfrak{g}_0)}  =
H^2(\mathfrak{g}_{-1}; (\mathfrak{g} _{-1},
\mathfrak{g}_0)_{*}).\eqno{(3)}
$$
The advantage of this reformulation: the Lie algebra cohomology (the
right hand side of (3)) is easier to compute (e.g., by means of the
package SupeLie when the general theory fails, or with the help of
various theorem).  At the same time the fine grading of Spencer
homology is not lost: the ${\mathbb Z}$-grading of $(\mathfrak{g}
_{-1}, \mathfrak{g}_0) _*$ which induces the grading $(3)$ of
$H^2(\mathfrak{g}_{-1}; (\mathfrak{g}_{-1}, \mathfrak{g}_0)_{*})$
coincides (up to a shift) with the oder of the structure functions.

{\bf Analogs of Weyl and Riemann tensors}.  Suppose $\mathfrak{g}_0$
contains a center (like in the case when a metric is preserved up to a
conformal factor).  Then the elements of $H^2(\mathfrak{g}_{-1};
(\mathfrak{g} _{-1}, \mathfrak{g}_0)_{*})$ are analogs of the Weyl
tensor.

Let $\hat{\mathfrak{g}}_0$ be the semisimple part of $\mathfrak{g}_0$
and let $\hat{\mathfrak{g}}_*$ be a shorthand for $(\mathfrak{g}_{-1},
\hat{\mathfrak{g}}_0) _*$.  The elements of $H^2 (\mathfrak{g}_{-1};
\hat{\mathfrak{g}}_*)$ are analogs of the Riemann tensor.

The relation between $\hat H=H^2 (\mathfrak{g}_{-1};
\hat{\mathfrak{g}}_*)$ and $H=H^2 (\mathfrak{g}_{-1};
(\mathfrak{g}_{-1}, \mathfrak{g}_0)_*)$ is more intricate for the
general $\hat{\mathfrak{g}}_{0}$ than in the Riemannian case
($\hat{\mathfrak{g}}_{0}=\mathfrak{o}(n)$) when $\hat H$ strictly
contains $H$.  In general, these spaces have common components
(conformally invariant, ``on shell'' ones) and have other components,
analogs of ``off shell'' components, cf.  \cite{Gro/Lei2}.

In the Riemann case, there are two ``off shell'' components: with the
highest weights $(2, 2)$ (the traceless Ricci tensor) and $(0, 0)$
(the scalar curvature).  Here the highest weights are given with
respect to the complexification $\mathfrak{L}=\mathfrak{sl}(2)_L\oplus
\mathfrak{sl}(2)_R$ of the $\mathfrak{o}(1, 3)$.  The Einstein equaton
(in vacum) is a vanishing condition of these components.  Remarkably,
there are no structure functions of lesser order.  If they had
existed, we would have to impose analogs of Wess-Zumino constraints to
be able to define the usual Riemann curvature tensor.

{\bf 4.  The description of $(\mathfrak{g}_{-1}, \mathfrak{g}_0)_*$
for the curved supergrassmannians}.  For the general curved
supergrassmannian of $(0, k)$-dimensional subsupermanifolds ${\cal S}$
in the $(0, n)$-dimensional supermanifold ${\cal T}$ let $\xi_1, ... 
, \xi_k$ be the coordinates of ${\cal S}$ and $\theta_1, ...,
\theta_{n-k}$ the remaining coordinates of ${\cal T}$.  Then setting
$\deg xi_i=0$ for all $i$ and $\deg \theta_j=1$ for all $j$ we get a
${\mathbb Z}$-grading of $\mathfrak{vect}(0|n)$ of the form
$$
\mathfrak{g}_0 =(\mathfrak{g}l(V)\otimes {\mathbb C} [\xi])\supplus
\mathfrak{vect}(\xi);\; \; \mathfrak{g}_{-1}=V\otimes {\mathbb C}
[\xi]; \eqno{(4)}
$$
where $V={\rm Span}(\frac{\partial}{\partial \theta _1}, ...  ,
\frac{\partial}{\partial \theta_{n-k}})$ is the identity
$\mathfrak{g}l(V)$-module, and $\supplus$ is the sign of a semidirect
sum of algebras: $\mathfrak{a}\supplus \mathfrak{b}$ with the ideal
$\mathfrak{a}$.

For $n=4$ we computed $H^2 (\mathfrak{g}_{-1}; (\mathfrak{g}_{-1},
\mathfrak{g}_0)_*)$ in the following cases:

(a) the general curved supergrassmannians;

(b) the supergrassmannians of subspaces through $0$, i.e., we removed
from $\mathfrak{vect}$ all partial derivatives (since this is not an
invariant formulation, it is better to say: we only considered the
vector fields that vanish at the origin);

(c) in case (b) we only considered volume-preserving transformations, 
i.e., we diminished $\mathfrak{g}_0$ as well:
$$
\mathfrak{g}_0=(\mathfrak{sl}(V)\otimes {\mathbb C} [\xi])\supplus
\mathfrak{sl}({\rm Span}(\xi));\quad \mathfrak{g}_{-1}=V\otimes
{\mathbb C} \xi.
$$
In particular, since $\mathfrak{g}_{-1}$ is isomorphic to the tangent
space at a point of the curved supergrassmannian, we see that its even
part in cases (a) -- (c) is the same $Gr_{2}^4$ while the tangent
space to the whole supermanifold at the ``origin'' is ${\rm
Span}(\xi_{i}\frac{\partial}{\partial \theta_{j}}: 1\leq i, j\leq 2$). 
So the number of odd coordinates of our model varies from 4 in case
(a) to 2 in cases (b) and (c).

{\bf Table}.  In the first line there are indicated the degrees, i.e.,
orders, of all nonzero structure functions and the rest of the table
lists their the weights (with respect to $\mathfrak{L}$) (superscript
denotes the multiplicity of the weight the subscript the degree of the
corresponding structure function).  The $\mathfrak{g}_0$-action is
nontrivial and glues distinct irreducible $(\mathfrak{g}_0)_{\bar
0}$-modules.  (We did not show the action though we have computed it.)

\begin{tabular}{|c|c|}
\hline
{\bf Odd structure functions}&{\bf Even structure functions}\cr
\hline
\begin{tabular}{|ccc|}
\hline
$-2$&$-1$&$0$\cr
\hline
$(11)$&$(01)^2$&$(11)$\cr
$(13)$&$(23) $& \cr
 &$(03) $& \cr
 &$(21) $& \cr
\hline
\end{tabular}&
\begin{tabular}{|ccc|}
\hline
$0$&$1$&$2$\cr
\hline
$(00)^2$&$(10)$&$(00)$\cr
$(02)$&$(12)$&$(02)$\cr
$(04)^2$&$(14)$&$(04)$\cr
$(22)$&$(32)$&$(22)$\cr
$(24)$& & \cr
$(40)$& & \cr
\hline
\end{tabular}\cr
\hline
\end{tabular}

The $(\mathfrak{g}_{0})_{\bar 0}$-modules whose highest weights are given
in the table are glued into $\mathfrak{g}_{0}$-modules as follows (an
arrow indicates a submodule).  The even tensors:
$$
\begin{array}{ccc}
(00)_{0}^2&\longrightarrow & (02)_{2};\cr
\searrow&(12)_{1}&\nearrow;\end{array}\quad
\begin{array}{ccc}(04)_{0}&\longrightarrow &(04)_{2};\cr
\searrow&(14)_{1}&\nearrow;\end{array}
$$
$$
(22)_{0}\longrightarrow(14)_{1}\longrightarrow(22)_{2}; \quad
(22)_{0}\longrightarrow(32)_{1}\longrightarrow(22)_{2}; 
$$
$$
(24)_{0}\longrightarrow(32)_{1}; \; (12)_{1}\longrightarrow(04)_{2}; \quad
(40)_{0}\longrightarrow(32)_{1}; \; (12)_{1}\longrightarrow(22)_{2}.
$$
The odd tensors:
$$
\begin{array}{l}
	(11)_{-2}\longrightarrow (23)_{-1};\; \;
	(01)_{-1}^2\longrightarrow(11)_{0};\cr
	(13)_{-2}\longrightarrow(23)_{-1}.\end{array}
$$

{\bf 5.  The Einstein equations}. The conventional EE in vacum are the 
conditions on the two tensors of degree 2 and weight $(00)$ and 
$(22)$, namely,
$$
R(22)=0\quad  {\rm  and }\quad R(00)=\lambda g, \eqno{(5)}
$$
where $\lambda\in {\mathbb C}$ is interepreted in terms of the
cosmological constant and $g$ is the metric preserved.

For an analog of the Einstein equations on the curved
supergrassmannian we may take the same vanishing conditions of the
2-nd order structure functions of weights $(00)$ and $(22)$ with
respect to $\mathfrak{L}$.  However, unlike the Einstein's case, we
have to vanish the constraints, the structure functions of lesser
orders, both even and odd.  The meaning of these analogs of
Wess-Zumino constraints is unclear to us.


\begin{thebibliography}{9}

\bibitem{del}
Deligne P. et al (eds.)  
\newblock {\it Quantum fields and strings: a course for
mathematicians}.  Vol.  1, 2.  Material from the Special Year on
Quantum Field Theory held at the Institute for Advanced Study,
Princeton, NJ, 1996--1997.  
\newblock AMS,
Providence, RI; Institute for Advanced Study (IAS), Princeton, NJ,
1999.  Vol.  1: xxii+723 pp.; Vol.  2: pp.  i--xxiv and 727--1501

\bibitem{Gro/Lei1}
Grozman P., Leites D., 
\newblock {\it {\em Mathematica}-aided study of Lie algebras 
and their cohomology.  From supergravity to ballbearings and magnetic 
hydrodynamics}
\newblock In: Ker\"anen V. (ed.)  {\em The second International 
Mathematica symposium}, Rovaniemi, 1997, 185--192
\bibitem{Gro/Lei2}
Grozman P., Leites D., 
\newblock {\it Supergravities and $N$-extended Minkowski 
superspaces for any $N$}.  
\newblock In: Wess J., Ivanov E. (eds.)  {\it 
Supersymmetries and quantum symmetries}. Proc.  
International Conference in memory of V.~Ogievetsky, June 1997, 
Lecture Notes in Physics 
\newblock {\bf 524},  Springer, 1999, 58--67
\bibitem{Lei/Ser/Vin}
Leites D., Serganova V., Vinel G., 
\newblock {\it Classical superspaces and related 
structures}.  
\newblock In: Bartocci C. et al.  (eds) {\em Differential Geometric 
Methods in Theoretical Physics}. Proc.  DGM-XIX, 1990, Springer, LN 
Phys.  
\newblock {\bf 375}, 1991, 286--297
\bibitem{Man}
Manin, Yu. 
\newblock {\it Gauge field theory and complex geometry}, 
\newblock Springer-Verlag, Berlin, 1997.  
\bibitem{Sal/Sez}
Salam A., Sezgin E., 
\newblock {\it Supergravities in diverse dimensions}, v.v. 
1, 2, World Scientific, 1989
\bibitem{Ste}
Sternberg S., 
\newblock {\it Lectures on differential geometry}, Chelsey, 2nd 
edition, 1985

\end{thebibliography}
\end{document}